# The Mechanism of Electrolyte Gating on High-$T_c$ Cuprates: The Role of Oxygen Migration and Electrostatics

Lingchao Zhang[†,‡], Shengwei Zeng*[,†,‡] , Xinmao Yin[‡,§,#], Teguh Citra Asmara[§], Ping Yang[§], Kun Han[†,‡], Yu Cao[†,‡], Wenxiong Zhou[†,‡], Dongyang Wan[†,‡], Chi Sin Tang[‡,§,∥], Andrivo Rusydi[†,‡,§], Ariando[†,‡,∥], Thirumalai Venkatesan*[,†,‡,∥,⊥,∇]

[†]NUSNNI-NanoCore, National University of Singapore, Singapore 117411
[‡]Department of Physics, National University of Singapore, Singapore 117551
[§]Singapore Synchrotron Light Source (SSLS), National University of Singapore, 5 Research Link, Singapore 117603
[#]SZU-NUS Collaborative Innovation Center for Optoelectronic Science & Technology, Key Laboratory of Optoelectronic Devices and Systems of Ministry of Education and Guangdong Province, College of Optoelectronic Engineering, Shenzhen University, Shenzhen, China 518060
[∥]NUS Graduate School for Integrative Sciences and Engineering (NGS), National University of Singapore, Singapore 117456
[⊥]Department of Electrical and Computer Engineering, National University of Singapore, Singapore 117576
[∇]Department of Materials Science and Engineering, National University of Singapore, Singapore 117575

*To whom correspondence should be addressed.
E-mail: nnizs@nus.edu.sg, venky@nus.edu.sg

**Abstract**

Electrolyte gating is widely used to induce large carrier density modulation on solid surfaces to explore various properties. Most of past works have attributed the charge modulation to electrostatic field effect. However, some recent reports have argued that the electrolyte gating effect in $VO_2$, $TiO_2$ and $SrTiO_3$ originated from field-induced oxygen vacancy formation. This gives rise to a controversy about the gating mechanism, and it is therefore vital to reveal the relationship between the role of electrolyte gating and the intrinsic properties of materials. Here, we report entirely different mechanisms of electrolyte gating on two high-$T_c$ cuprates, $NdBa_2Cu_3O_{7-\delta}$ (NBCO) and $Pr_{2-x}Ce_xCuO_4$ (PCCO), with different crystal structures. We show that field-induced oxygen vacancy formation in CuO chains of NBCO plays the dominant role while it is mainly an electrostatic field effect in the case of PCCO. The possible reason is that NBCO has mobile oxygen in CuO chains while PCCO does not. Our study helps clarify the controversy relating to the mechanism of electrolyte gating, leading to a better understanding of the role of oxygen electro migration which is very material specific.

The electronic properties of strongly-correlated materials depend strongly on carrier density and experiments where the carrier density is modified systematically are improtant for understanding the fundamental physics behind the studied material phenomena.[1] Electric field effect induced charge modulation is preferred to chemical doping because the carrier density can be tuned quasi-continuously without introducing chemical or structural changes.[1-10] However, the electric field effect on the strongly-correlated materials remains small in the field effect transistor (FET) devices using conventional solid dielectric layers. This is because most of the phenomena occur at carrier densities exceeding $10^{14}$ cm$^{-2}$, which is not easy to achieve without extremely high K layers. For instance, only several Kelvin of $T_c$ shift was realized for high-$T_c$ cuprates using solid dielectrics.[2-7]

The situation was changed recently when the technique of electrolyte gating using ionic liquids (ILs) as gate dielectrics was introduced.[11-13] Based on the principle of electrochemical capacitor, ions in the ILs are separated and accumulated on the solid surface of both electrodes under gate voltages. Meanwhile, opposite charges of equivalent density accumulate on the electrode side, creating an electric double-layer (EDL) effectively working as an interface capacitor.[14] The nanoscale separation of the EDL results in ultrahigh electric field of the order of 10 MV cm$^{-1}$ with induced carrier density up to $10^{15}$ cm$^{-2}$. This technique has been used to induce insulator-to-metal transition,[13, 15-17] superconductivity[18-22] and other electronic phase transitions.[23-27] Continuous superconductor-insulator transition (SIT) has also been obtained in high-$T_c$ cuprates.[28-32] The charge modulation mechanisms in these works were attributed to electrostatic field effect. However, several reports recently argued that the electrolyte gating effect originated from field-induced oxygen migration,[33-37] hydrogen injection[37-40] and structure modification.[41-43] Indeed, oxygen migration can have a significant influence on the properties of a broad spectrum of materials.[44] Hence, both electrostatic charging and oxygen migration could induce large change of carrier density and result in similar phenomena, giving rise to a controversy about the gating mechanism. In order to identify the gating effect and also explore potential device application in a specific specimen, it is therefore vital to relate the role of electrolyte gating to the intrinsic properties of materials. Here, we reveal the link between electrolyte gating mechanism and the intrinsic crystal structure by comprehensive comparison of two types of high-$T_c$ cuprates, RE-123 and 214-type, represented by $NdBa_2Cu_3O_{7-\delta}$ (NBCO) and $Pr_{2-x}Ce_xCuO_{4-\delta}$ (PCCO) respectively (unit cells shown in Figure S1), through the measurements of electrical transport, spectroscopic ellipsometry and X-ray absorption.

**RESULTS AND DISCUSSION**

Figure 1a shows the schematic diagram of an electric double-layer transistor (EDLT) device which consists of a Hall-bar channel and a lateral gate electrode (see Methods). In both NBCO and PCCO devices, positive gate voltages $V_g$ were applied, the mobile cations accumulated on the transport channel and induced electrons on the surface of the thin films. The leakage currents are within $2\times10^{-8}$ A for $V_g$ ranging from 0 to 1.8 V for NBCO and 0 to 2 V for PCCO (Supporting Information Figure S4). These negligible small leakage currents indicate good performance of EDLT devices. Figure 1b and 1c are the on and off characterization of electrolyte gating on NBCO and PCCO films, respectively. The sheet resistance $R_s$ was recorded as a function of time $t$. The $V_g$ was switched on at a set voltage and kept on for a suitable duration, then was switched off to 0 V and kept off for another similar duration. Each curve for NBCO can be divided into four regions (A, B, C and D), taking the case of ON $V_g$ = 1 V as an example. In the gate voltage ON region, as the $V_g$ is applied, the $R_s$ shows an instantaneous increase in a short duration (region A) followed by a linear increase with time under a constant $V_g$ (region B). In the gate voltage OFF region, as the $V_g$ was switched off, the $R_s$ instantly decreases slightly (region C) and becomes constant with time (region D). Even though the $V_g$ was kept off for more than 30 min, the $R_s$ did not reduce back to the original value. This means that the electrostatic charging and depleting, in which the sample should recover its original state after the $V_g$ is switched off, is not the main origin of the gating effect in NBCO. These results suggest that oxygen electromigration dominates the transport properties during the gating process in NBCO EDLT. After the application of a constant $V_g$, oxygen vacancies were continuously created with time (electrons were introduced), and therefore, the $R_s$ increases linearly in region B. After removal of $V_g$, since the oxygen migration is absent and the oxygen vacancies are still present in NBCO, the $R_s$ remains constant in region D. We also performed the gating induced on and off characterizations on NBCO at various $N_2/O_2$ gas flow ratios, and it is showed that the irreversibility as the $V_g$ is switched off is present at all different gas flow ratios (Supporting Information Section 4 and Figure S6). This suggested that oxygen vacancies are prone to be created in NBCO under the influence of electric field, even in pure $O_2$ gas atmosphere.

In the linear region B, the slope of $R_s$, $dR_s/dt$, was obtained by fitting the data and plotted as a function of $V_g$ in Figure 1d. One can see that $dR_s/dt$ increases with increasing $V_g$, suggesting that oxygen migration is more efficient at higher $V_g$. Moreover, the $dR_s/dt$ versus $V_g$ data can be well fitted by a line in the log-linear scale. This supports the model of field-induced oxygen migration expected to depend exponentially on the electric field which affects the diffusion coefficient of oxygen migration between sample film and IL during electrolyte gating.

Interestingly, for PCCO, the on and off characteristics show totally different behaviour, as shown in Figure 1c. The $R_s$ decreases sharply as a certain $V_g$ is applied (electrons were introduced), and immediately tends to a saturation value which is proportional to the $V_g$ with a long time constant. As the $V_g$ is switched off, the $R_s$ immediately increases and then slowly recovers to the original value with a similar time constant. The slope $dR_s/dt$ for PCCO obtained from the approximately linear region (in the region where the dash line is drew) is also plotted as a function of $V_g$ in Figure 1d. The slope only shows a slight increase with increasing $V_g$, compared with that of NBCO, suggesting that the $R_s$ of PCCO saturates with time at a similar rate regardless of the value of $V_g$. These behaviours indicate a capacitor-like charging mechanism in PCCO EDLT and the time constant may be associated with the ion accumulation dynamics in the IL. Hence, the mechanism of electrolyte gating on PCCO is mainly electrostatic field effect. The gating induced on and off characterizations at various $N_2/O_2$ gas flow ratios were also performed, and it is showed that the reversibility as the $V_g$ is switched off is present at all different gas flow ratios (Supporting Information Section 4 and Figure S6), further suggesting the electrostatic effect in PCCO.

Figure 2a shows the sheet resistance-temperature (RT) curves of 7-uc NBCO film before gating, gating at various $V_g$ and after gating. One can see that the as-grown NBCO film is superconducting (RT0) and it can be tuned to insulating state at $V_g$ = 1.4 and 1.6 V (RT2 and RT3). The $V_g$ was then switched off to 0 V, the sample was heated to 300 K and the IL was removed by acetone cleaning quickly and blown dry with nitrogen gas. After the removal of IL, the sample was still insulating and shows higher $R_s$ (RT4). This means that the oxygen vacancies which were created during the gating process, were still present in NBCO after the removal of IL. It has been reported that in RE-123 cuprates, reduction of oxygen in the Cu-O chain causes a decrease in hole concentration and induce superconductor to insulator transition.[45-46] Therefore, the induced insulating state of NBCO after gating may be due to the decrease in oxygen composition in Cu-O chain during the gating process. Note that the quasi-continuous superconductor to insulator transition near the pair quantum resistance $R_Q = h/(2e)^2$ = 6.45 kΩ could also be tuned by gated induced oxygen reduction (Supporting Information Fig. S5). After gating, the sample was oxygen annealed at 540 °C for 30 min, and it shows significantly reduced $R_s$ and recovery of superconducting transition, even though it could not recover to the initial state (RT5). This suggests that the chemical composition in the film was probably altered by electrolyte gating, making the gating effect irreversible after the removal of IL which will be discussed in the following text. Note that the as-grown films have been oxygen annealed, ensuring full oxygenation. The $R_s$ reduction and recovery of superconducting transition by oxygen annealing clearly reflects oxygen vacancy formation in the film during gating.

Figure 2b shows the $R_s$-$T$ curves for PCCO EDLT before and after gating and $R_s$-$T$ curves at various $V_g$. Upon increasing the $V_g$, electrons were introduced and insulating to superconducting phase transition was observed (inset of Figure 2b). In contrast to NBCO, after electrolyte gating and removing the IL, the initial insulating state of PCCO was recovered, which can be seen from the fact that the $R_s$-$T$ curves before and after gating show reasonable agreement. This result indicates that electrostatic charging is the main origin of electrolyte gating in PCCO.[31]

To further confirm the formation of oxygen vacancies in NBCO and identify the specific lattice location, we conducted spectroscopic ellipsometry measurements. The motivation is based on a previous study on YBa$_2$Cu$_3$O$_{7-\delta}$ (YBCO) that the peaks of 4.1 eV and 1.7 eV in the dielectric function become very strong and sharp for oxygen deficient samples.[47] Particularly, the 4.1 eV peak is identified with the existence of Cu$^{1+}$ (oxygen deficiency) in a highly localized O-Cu(1)-O complex (mainly the CuO chain). Figure 3a shows the real part $\varepsilon_1$ and imaginary part $\varepsilon_2$ of dielectric function $\varepsilon_1 + i\varepsilon_2$ for the as-grown oxygen deficient and oxygen annealed NBCO films. The oxygen deficient film was in-situ post-annealed in vacuum (< 0.1 mTorr) at 520 °C for 20 min during the cooling procedure. After the ellipsometry measurement, the oxygen deficient sample was transferred into the PLD chamber for oxygen annealing at 540°C and 600 Torr for 1 hour, ensuring it is fully oxygenated. One can see that the oxygen deficient NBCO film shows strong peaks at 4.1 eV and 1.7 eV, consistent with previous results in YBCO.[47] Figure 3b shows the ellipsometry measurements on NBCO film before and after electrolyte gating.

We can see that the main features of the change of dielectric function by electrolyte gating are the same as those for the oxygen deficient sample in Figure 3a. Particularly, the 4.1 eV peak increased significantly after electrolyte gating, indicating oxygen vacancy formation in CuO chains. The decrease of the Drude response at low-energy region after gating suggested the decrease of conductivity, consistent with electrical transport measurements. In contrast, for PCCO, the measured dielectric function was not affected by electrolyte gating as shown in Figure 3c, indicating that the gated-induced oxygen vacancies are absent. Therefore, we can conclude that the dominant mechanism of electrolyte gating on NBCO is field-induced oxygen migration out of CuO chains while the mechanism on PCCO is mainly electrostatic field effect.

Figure 4 shows the in-plane X-ray absorption spectroscopy (XAS) spectra on Cu $L$-edge and O $K$-edge of NBCO film before and after electrolyte gating.[48] It can be seen that in the Cu $L_3$-edge spectra, the satellite ligand-hole peak (labeled B) besides the white line peak (labeled A) decreases after gating, indicating the decrease of carrier density. In the O $K$-edge spectra, the prepeak (labeled C) which is ascribed to the Zhang-Rice singlet state[49] and is related to doped holes in $CuO_2$ planes decreases largely after gating.[50-51] The hole depletion in $CuO_2$ planes implies the occurrence of charge transfer from CuO chains to $CuO_2$ planes. A previous XAS study reported that the electric field doping of $CuO_2$ planes in NBCO and YBCO was indirect, which is consistent with our results.[32, 52]

In RE-123 cuprate system, there are CuO chains and the structure and physical properties depend considerably on the oxygen content in CuO chains. The oxygen in CuO chains are prone to reduce from 7 to 6 under thermal treatment at moderate temperature of 450 ~ 550 °C under low oxygen partial pressure. With decreasing oxygen composition, the unit cell expands, the hole density in $CuO_2$ plane decreases, and therefore, the sample changes from superconductor to insulator.[45-46] The X-ray diffraction was also performed on NBCO film before and after gating, and the data showed an expansion of the unit cell after gating (Supporting Information Section 6 and Figure S9). This coupled with spectroscopic data and the gated induced SIT indicate that the electrolyte gating causes the reduction of oxygen in CuO chain, and therefore, the reduction of hole density in $CuO_2$ plane. However, in 214-type cuprate system, CuO chain is absent. Even though reduction is required to obtain superconductivity, the oxygen content is within 3.9 ~ 4 regardless of the doping level[53] and the lattice constant is almost unchanged under reduction even at temperature above 1000 °C, suggesting that PCCO is stable under thermal treatment at low oxygen partial pressure.[53] Furthermore, the film in the current study is underdoped $Pr_{1.9}Ce_{0.1}CuO_4$, which is not superconducting even if the oxygen is reduced.[54] However, superconductivity in PCCO-EDLT is obtained by liquid gating. Therefore, it is difficult to create oxygen vacancies by electrolyte gating, and the observed insulator to superconductor transition is the result of electrostatic charging. In another 214-type hole-doped cupate $La_{2-x}Sr_xCuO_4$, the initially insulating underdoped film could also be induced to be superconducting by electrolyte gating.[28] Since in $La_{2-x}Sr_xCuO_4$ oxygen annealing is required to obtain superconductivity, oxygen vacancies are absent in such liquid-gated superconductor. Therefore, these results indicate that in liquid-gated cuprates, oxygen vacancies are prone to be created in cuprates with CuO chains in which the oxygen is known to be loosely bound and mobile; while in cuprates with only $CuO_2$ plane, electrostatic charging dominates.

It has been reported that electrolyte-gated oxygen vacancy creation caused the increase in the surface roughness in crystalline $TiO_2$ and $SrTiO_3$.[34, 55] For NBCO in the current study, it was found that the optical surface roughness (from ellipsometry model fitting) increased significantly after gating, from 2.5 nm to 5 nm. Electrostatic field effect generally takes effect within only ~ 2 nm (1 or 2 uc) from the surface because it is limited by the electrostatic screening length,[28] but oxygen migration does not have such depth limitation. This point also supports the occurrence of oxygen migration out of thin film. The increase in surface roughness after gating could also been seen from the atomic force microscopy (AFM) images (Supporting Information Section 6 and Figure S10).

Chemical reaction on electrolyte-gated oxide surfaces causing irreversible deterioration effects on the electrical properties has been observed.[34, 55] In Figure 2a, after the removal of IL, the RT curve (RT4) shows higher resistance, suggesting the presence of chemical reaction. Since the gating is performed at lower temperature of 230 K at which the IL condenses into a rubberlike state and most chemical reactions are suppressed,[28-29] the chemical reaction may occur at higher temperature in the cooling or warming process. In fact, as the IL was put on the film surface, the room-temperature resistance could increase continuously with time even without application of $V_g$ and the initially superconducting sample could be damaged to become an insulator (Supporting Information Section 5 and Figure S7b). The key take away here is that while working with materials like NBCO which could react with the IL at room temperature it is prudent to get the samples to temperatures below 240 K as soon as possible after the IL is put on the sample surface. Once such rapid transfer can take place, the chemical reaction is mitigated and the superconducting transition is preserved (with minimal deterioration). We also did control experiments of ellipsometry to confirm that the change of dielectric function and thus the oxygen vacancy formation in CuO chains are not caused by the chemical reaction with IL (Supporting Information Figure S8). Therefore, the oxygen migration is caused by liquid gating at 230 K when the $V_g$ is applied, while chemical reaction occurs at higher temperature when IL is covered. Due to the presence of chemical reaction damage, even though gate-induced oxygen vacancies were recovered by oxygen annealing, the resistance could not recover its initial state, as shown in RT5 in Figure 2a.

**CONCLUSION**

In conclusion, we find that the mechanism of electrolyte gating is totally different for two high-$T_c$ cuprates. In NBCO with loosely bound CuO chain, electrolyte gating caused the oxygen reduction in CuO chain and a subsequent hole depletion in $CuO_2$ plane which finally leads to superconductor to insulator transition. In contrast, the electron charging induced insulator to superconductor transition in PCCO is mainly dominated by electrostatic field effect. These results suggest that the charging and depletion arising from liquid electrolyte gating is very dependent on the material studied and strongly influenced by how tightly the oxygen is bound to the system. The current study sheds light on the mechanism of electrolyte gating, which could be a guide for future work and may also help develop the technique of electrolyte gating into functional devices.[56-57]

**METHODS**

NdBa$_2$Cu$_3$O$_{7-\delta}$ (NBCO) thin films were grown on single crystalline TiO$_2$-terminated SrTiO$_3$ (STO) (001) substrates by pulsed laser deposition (PLD) using a ceramic NBCO target. The deposition temperature is 700 °C and oxygen partial pressure $P_{O2}$ is 50 mTorr. After deposition, the samples were in-situ post-annealed at 520 °C and $P_{O2}$ of 600 Torr for 20 min in the chamber during the cooling procedure. During the deposition, an in situ reflection high energy electron diffraction (RHEED) was used to monitor the thickness. For the gated thin film, the thickness of NBCO is 7 unit cell (uc). Before deposition of NBCO, the treated STO substrate was patterned into Hall-bar configuration by conventional photolithography and depositing an amorphous AlN layer as hard mask. After NBCO growth, we did not carry out any processing step such as lithography because exposing the samples to chemicals will deteriorate the quality of the surface. During the electrical transport measurement, the electrodes were then bonded with Al wire and a droplet of silver paint was put on each bonding place to ensure ohmic contact and prevent peel-off during cooling. For the PCCO sample, nominal 1-uc underdoped Pr$_{1.9}$Ce$_{0.1}$CuO$_4$ on 4-uc undoped Pr$_2$CuO$_4$ (PCO) were grown on TiO$_2$-terminated STO (001) substrates by PLD. The thin films were deposited at 790 °C under $P_{O2}$ of 0.25 mbar and then cooled down to room temperature from 720 °C in vacuum ($P_{O2} < 10^{-4}$ mbar) at a cooling rate of 20 °C/min. The patterning of PCCO device is similar to NBCO. Since the PCCO films are stable in chemicals in lithography process, we performed a second-time photolithography to deposit Cr/Au (10/70nm) electrodes. The detailed thin film growth and device fabrication are seen in Supporting Information Section 1.

The IL used is DEME-TFSI (N, N-diethyl-N-(2-methoxyethyl)-N-methylammonium bis-trifluoromethylsulfonyl-imide) from Kanto Chemical Co. After covering the IL, the sample was then placed in the chamber of a Quantum Design Physical Property Measurement System (PPMS) for the transport measurement. The gate voltage $V_g$ was applied and changed at 230 K. During cooling for the resistance-temperature (RT) measurements, the $V_g$ was kept constant. The ellipsometry measurements were performed by Woollam V-VASE spectroscopic ellipsometer with a spectral range of 0.6−6.5 eV. The X-ray absorption spectroscopy (XAS) measurements were conducted from the SINS beamline at Singapore Synchrotron Light Source (SSLS) using linearly polarized beam *via* a total electron yield mode. The detailed measurement process is seen in Supporting Information Section 2.

## ASSOCIATED CONTENT

**Supporting Information**

The Supporting Information is available free of charge on the ACS Publications website.

Detailed thin film growth, device fabrication and measurement methods, continuous modulation of SITs in NBCO by electrolyte gating, time-dependence resistance for both NBCO and PCCO at various $N_2/O_2$ gas flow ratios, control experiments of transport and ellipsometry measurements, and XRD and AFM measurement data for NBCO.

Conflict of Interest: The authors declare no competing financial interests.

ACKNOWLEDGEMENTS

We thank W. M. Lv, Z. Huang, C. J. Li, H. J. H. Ma and R. S. Nagarajan for their useful help and discussions. We would like to acknowledge the support from NRF-CRP15-2015-01 (Oxide electronics on silicon beyond Moore) and the NUS FRC (AcRF Tier 1 grant no. R-144-000-346-112, R-144-000-364-112, R-144-000-368-112). We would also like to acknowledge the Singapore Synchrotron Light Source (SSLS) for providing the facility necessary for conducting the research. The Laboratory is a National Research Infrastructure under the National Research Foundation Singapore. P. Yang is supported by SSLS *via* NUS Core Support C-380-003-003-001.

**Figures Legends**

**Figure 1.** Schematic NBCO EDLT and the gate voltage-dependent resistances. (a) The schematic diagram of a NBCO EDLT device. The width of the Hall-bar channel is 100 μm and the distance between two voltage probes is 160 μm. The on and off characterizations of electrolyte gating on NBCO (**b**) and PCCO films (**c**). The ON gate voltage $V_g$ is varied from 0 to 1 V in steps of 0.2 V. (**b**) For NBCO film, the $V_g$ is switched on and kept on for about 30 min, and then switched off and kept off for about 30 min. For the sake of discussion, the data for ON $V_g = 1$ V has been divided in to four distinct regions: A, B, C and D. (**c**) For PCCO film, the $V_g$ is switched on and kept on for about 60 min, and then switched off and kept off for about 60 min. The dashed lines are used to illustrate the approximately linear behaviour. (**d**) The slope in the region B for NBCO and in the approximately linear region for PCCO are extracted and plotted as a function of ON $V_g$. In both (**b**) and (**c**), the gating experiments were performed at 230 K.

**Figure 2.** Gating effect on sheet resistance-temperature (RT) curves. (**a**) RT curves of NBCO film for as-grown state (RT0), gating at various $V_g$ (RT1 to RT3), after gating and removal of IL by acetone (RT4), and oxygen annealing (RT5). (**b**) RT curves of PCCO film before and after gating, showing the full recovery of electrolyte gating effect and no visible deterioration effects due to IL. The inset is the modulation of PCCO film from insulating state to superconducting state by electrolyte gating.

**Figure 3.** Spectroscopic ellipsometry measurement of dielectric function $\varepsilon_1 + i\varepsilon_2$ for NBCO and PCCO films. (**a**) Comparison of the oxygen deficient NBCO film (by post-annealing in vacuum) with the film fully oxygenated (by oxygen annealing at 540 °C for 1 hour). **(b)** Comparison of the as-grown fully oxygenated NBCO film (before gating) with the film electrolyte gated at 230 K followed by removal of IL by acetone (after gating). **(c)** Comparison of as-grown PCCO film (before gating) with the film electrolyte gated at 230 K followed by removal of ionic liquid by acetone (after gating). In each figure, the top panel is for real part $\varepsilon_1$ and bottom panel is for imaginary part $\varepsilon_2$.

**Figure 4.** X-ray absorption spectroscopy before (as-grown) and after (electrolyte gated and removal of IL) gating on NBCO film, showing charge transfer from the chains to the planes. **(a)** The Cu $L_3$-edge spectra. Two peaks are indicated by "A" (the white line peak) and "B" (the ligand-hole peak). The decrease of peak B indicates decrease in carrier density. **(b)** The O K-edge spectra. The prepeak of Zhang-Rice peak ("C") which is related to doped holes decreases largely after gating.

**FIGURES**

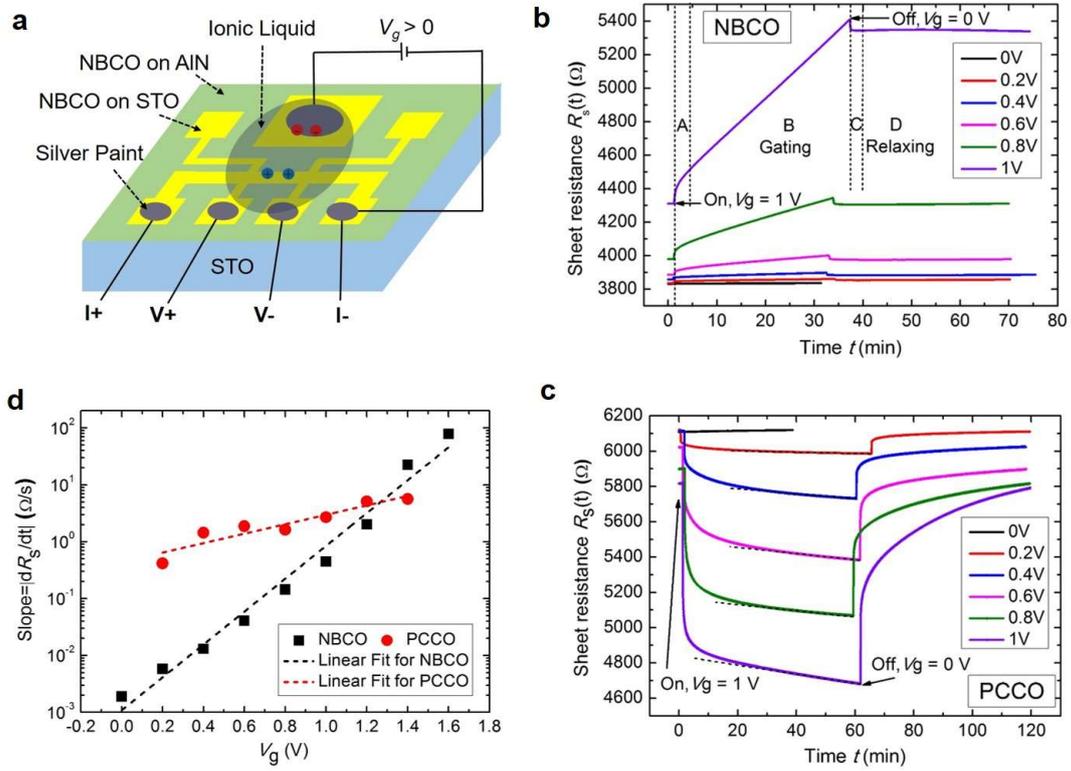

Figure 1

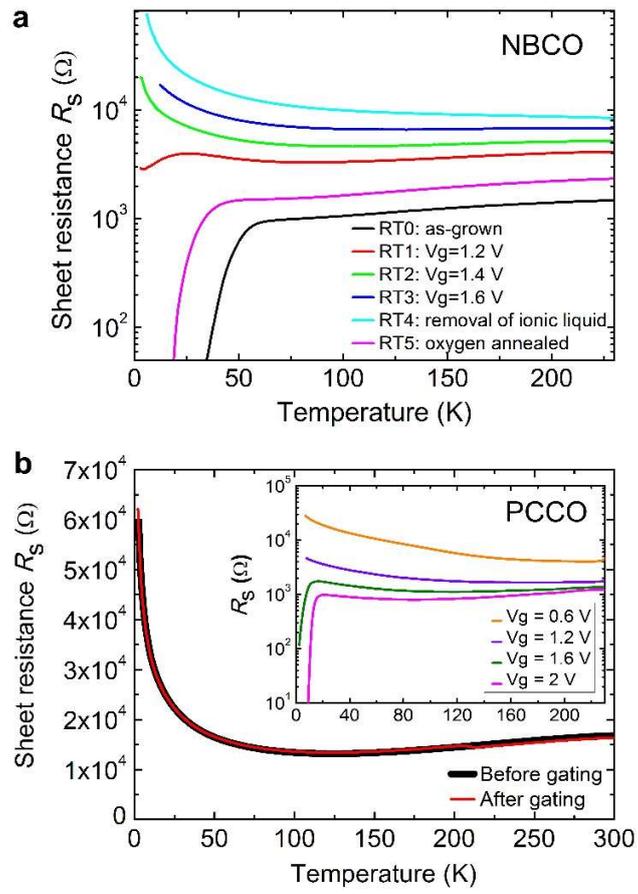

Figure 2

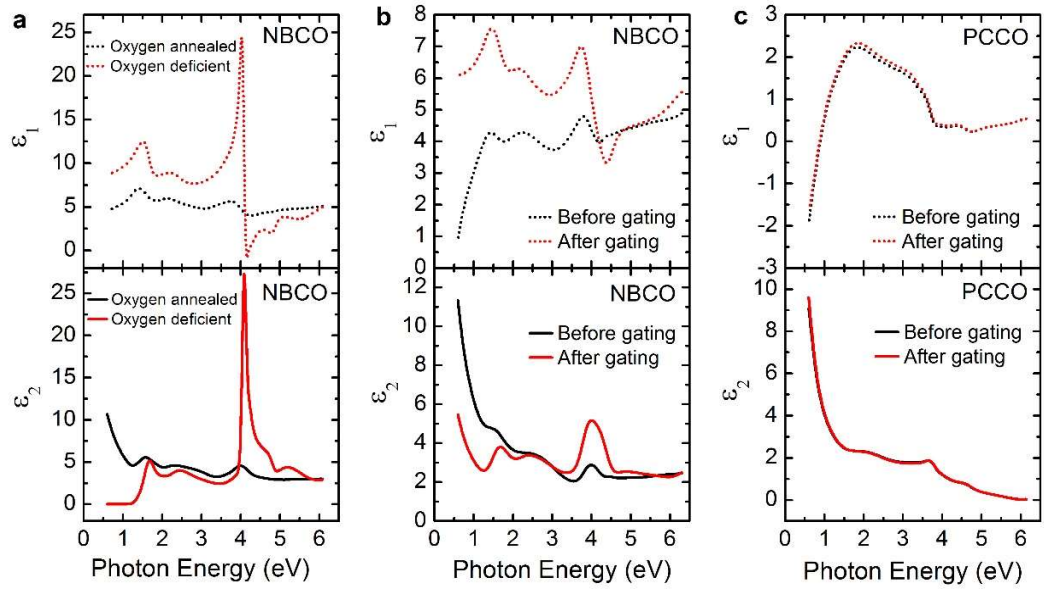

Figure 3

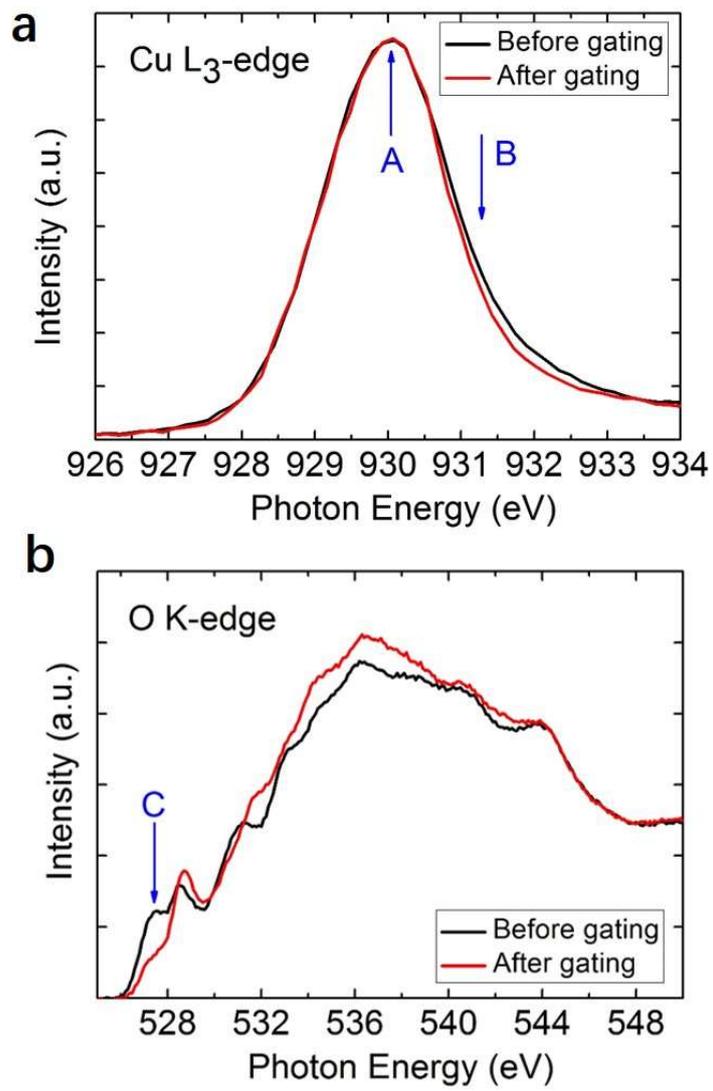

Figure 4

# Supporting Information for

# The Mechanism of Electrolyte Gating on High-$T_c$ Cuprates: The Role of Oxygen Migration and Electrostatics

Lingchao Zhang[†,‡], Shengwei Zeng*[†,‡], Xinmao Yin[‡,§,#], Teguh Citra Asmara[§], Ping Yang[§], Kun Han[†,‡], Yu Cao[†,‡], Wenxiong Zhou[†,‡], Dongyang Wan[†,‡], Chi Sin Tang[‡,§,∥], Andrivo Rusydi[†,‡,§], Ariando[†,‡,∥], Thirumalai Venkatesan*[†,‡,∥,⊥,∇]

[†]NUSNNI-NanoCore, National University of Singapore, Singapore 117411
[‡]Department of Physics, National University of Singapore, Singapore 117551
[§]Singapore Synchrotron Light Source (SSLS), National University of Singapore, 5 Research Link, Singapore 117603
[#]SZU-NUS Collaborative Innovation Center for Optoelectronic Science & Technology, Key Laboratory of Optoelectronic Devices and Systems of Ministry of Education and Guangdong Province, College of Optoelectronic Engineering, Shenzhen University, Shenzhen, China 518060
[∥]NUS Graduate School for Integrative Sciences and Engineering (NGS), National University of Singapore, Singapore 117456
[⊥]Department of Electrical and Computer Engineering, National University of Singapore, Singapore 117576
[∇]Department of Materials Science and Engineering, National University of Singapore, Singapore 117575

*To whom correspondence should be addressed.
E-mail: nnizs@nus.edu.sg, venky@nus.edu.sg

## 1. Thin film growth and device fabrication

$NdBa_2Cu_3O_{7-\delta}$ (NBCO) thin films were grown on single crystalline $SrTiO_3$ (STO) (001) substrates by pulsed laser deposition (PLD) using a ceramic NBCO target. The unit cell of NBCO is sketched in Figure S1a. STO substrate was pre-treated with HF ultrasonic etch for 30 seconds followed by annealing for 1.5 h at 950 °C. An optimized superconducting critical temperature $T_{c0}$ of 93 K was obtained for deposition temperature of 700 °C and deposition oxygen pressure $P_{O2}$ of 50 mTorr for films thicker than 50 nm. The laser energy density on the target was fixed at about 1.8 J/cm$^2$ and the laser frequency was set as 2 Hz. The samples were in-situ post-annealed at 520 °C and $P_{O2}$ of 600 Torr for 20 min in the chamber during the cooling procedure. The cooling rate was kept at 10 °C/min. In-situ reflection high energy electron diffraction (RHEED) oscillations were observed for the first 13-unit cell (uc) growth, reflecting a Stranski-Krastanov growth mode. Figure S2 shows the RHEED patterns before and after film growth and the RHEED intensity oscillations for the initial 13-uc NBCO growth. One oscillation represents one unit-cell growth, as confirmed by X-ray reflectivity (XRR) measurement. This observation allows us real-time control of the film thickness during deposition in an accurate way. The film with thickness of 7 uc was found to be best to realize the modulation of superconductor-insulator transition (SIT) by electrolyte gating. All the NBCO samples used in this paper except the one grown as oxygen-deficient sample were prepared under the same conditions and showed high reproducibility. The only difference for the oxygen-deficient sample was that it was in-situ post-annealed in vacuum (< 0.1 mTorr) instead of in oxygen (600 Torr) at 520 °C for 20 min during the cooling procedure.

Figure S3 shows the schematic overview of the fabrication processes of electrolyte gating devices. The treated STO substrate was patterned into Hall-bar configuration by photolithography and depositing an amorphous AlN layer as hard mask. The AlN layer was deposited by PLD at room temperature and had a thickness of about 200 nm. After patterning, the substrate surface was still clean and atomically flat with $TiO_2$ termination. The patterned STO was then used for NBCO thin film deposition. The NBCO film grown on AlN layer was highly insulating during the whole experiments, as monitored by transport measurements. This ensured the leakage current during electrolyte gating to be negligibly small (Figure S4). The electrodes were then bonded with Al wire and a droplet of silver paint was put on each bonding place to ensure ohmic contact and prevent peel-off during cooling. A second-time photolithography and Cr/Au electrode deposition were circumvented to prevent damage to NBCO film. The samples for electrical transport measurements were prepared on patterned STO substrates and made to "electrolyte gating devices" using this method. While the samples for optical measurements (ellipsometry, XAS, XRD) were prepared on bare STO substrates and made to "simple gating devices" by simply scratching a line on as-grown films using a diamond scribe.

The details for thin film growth of $Pr_{2-x}Ce_xCuO_4$ (PCCO) can be found in literature.[1] The unit cell of PCCO is sketched in Figure S1b. A buffer layer of 4-uc $Pr_2CuO_4$ (PCO) together with the film of 1-uc underdoped $Pr_{1.9}Ce_{0.1}CuO_4$ were grown at 790 °C under $P_{O2}$ of 0.25 mbar by PLD. The cooling from 720 °C to room temperature was in vacuum ($P_{O2} < 10^{-4}$ mbar). The fabrication processes of electrolyte gating devices for PCCO are similar to NBCO. In contrast to NBCO film, the electrical properties of PCCO film are quite stable, enabling us to do a second-time photolithography to deposit Cr/Au (10/70nm) electrodes. For better comparison and easier model fitting, the PCCO sample for ellipsometry measurements was grown without PCO buffer layer but directly with 7-uc PCCO film. The as-grown PCCO/STO films were made in to simple gating devices by simply scratching a line by a diamond scribe similar to the case of NBCO.

2. Measurement methods

The ionic liquid used is DEME-TFSI (N, N-diethyl-N-(2-methoxyethyl)-N-methylammonium bis-trifluoromethylsulfonyl-imide) from Kanto Chemical Co. The ionic liquid condenses into a rubbery state at 240 K and changes to a glassy state at 190 K. Between 240 K and 190 K, most chemical reactions are suppressed while ions in the ionic liquid are still mobile. Below 190 K, all the ions are frozen.[2-4]

For electrical transport measurements, after putting the ionic liquid over the film channel and gate pad, the device was immediately transferred into PPMS chamber, pumped to vacuum and cooled down to 230 K. Gate voltage $V_g$ was applied and changed at 230 K. During cooling for the resistance-temperature (RT) measurements, the $V_g$ was kept constant. Figure S4 shows the leakage current as a function of $V_g$. The leakage current is within $2\times10^{-8}$ A for gate voltage ranging from 0 to 1.8 V for NBCO and 0 to 2 V for PCCO. The gate voltages of 1.8 V for NBCO and 2 V for PCCO are enough to induce superconductor-insulator transitions (Fig. 2 in the main text). These negligible small leakage currents indicate good performance of EDLT devices.

For optical measurements, as-grown films on bare STO substrates were used as samples before gating. Then the samples were made to simple gating devices because a larger area of sample film was needed for optical measurements. The scratched line separated the sample film into two parts. Linear four-point-probe (4PP) was employed for wire bonding on the larger part. Silver paint was put on each bonding place to ensure ohmic contact as well as on the smaller part to make a large gate pad. After putting the ionic liquid over the whole device, it was immediately transferred into PPMS chamber, pumped to vacuum and cooled down to 230 K. The 4PP resistance was then tuned by electrolyte gating at 230 K and the sample was induced to be superconductor-insulator transitions. Then the gate voltage was switched off, the sample was warmed up, the ionic liquid was removed by acetone, and the sample was blown dry with nitrogen gas, to obtain the gated samples for optical measurements. After the $V_g$ was switched off, the sample charge will undergo some relaxation and the resistance will reduce to some extent. Hence, the heating and removal procedures should be fast enough before full relaxation. Fortunately, the relaxation was not so fast for NBCO case, enabling us to detect the residual electrolyte gating effect in the gated samples.

The ellipsometry measurements were performed by Woollam V-VASE spectroscopic ellipsometer with a spectral range of 0.6−6.5 eV. The X-ray absorption spectroscopy (XAS) measurements were conducted from the SINS beamline at Singapore Synchrotron Light Source (SSLS) using linearly polarized beam *via* a total electron yield mode. The X-ray diffraction (XRD) measurements were conducted from the XDD beamline at SSLS. The atomic force microscopy (AFM) measurements were performed by Agilent 5500 AFM in tapping mode.

3. Continuous modulation of SIT in NBCO by electrolyte gating

Figure S5a presented continuous modulation of SIT in NBCO by electrolyte gating, including more than twenty resistance-temperature (RT) curves. By finding and connecting those locally lowest or highest $R_s$ points, we can distinguish three regional boundaries and four electronic regions: metal, insulator, superconductor (SC) and low temperature ($T$) upturn. The use of those points is physical and reasonable because they are essentially the critical points where one phase fades away and another emerges. For example, the points in the insulator-SC boundary are the starting points of superconducting (SC) transition, because it is the emergence of local SC grains that causes the resistance to reduce with decreasing temperatures. Since the transition width of SC transition should vanish for the bulk, we thus can use the temperatures in the insulator-SC boundary to mimic the bulk $T_c$. That is to say, the insulator-SC boundary can be regarded as the phase boundary of the SIT physically, by ignoring the influence from the form of granular thin film. Figure S5b shows the enlarged region for the insulator-SC boundary and low $T$ upturn boundary. The two boundaries become converged at a point of 6126 Ω. What's more, we tried to exclude the influence from finite $T$ and the low $T$ upturn behaviour. On one hand, since not all high-$T_c$ cuprates have the low *T* upturn behaviour, we thus can ignore this behaviour when focusing on the study of SIT. On the other hand, physically, quantum phase transitions (QPTs) should occur at absolute 0 K. We thus extend the insulator-SC boundary to 0 K, resulting in a point at around 6450 Ω. This point should be related to a quantum critical point (QCP) and the critical resistance. Interestingly, the value is the pair quantum resistance $R_Q = h/(2e)^2$, suggesting that the SIT in NBCO is a QPT driven by quantum phase fluctuation and Cooper pair localization.[3-4]

## 4. Time-dependence resistance at various $N_2/O_2$ gas flow ratios

Figure S6 shows sheet resistance $R_s$ as a function of time at various $N_2/O_2$ gas flow ratios. For NBCO, at all different gas flow ratios, the resistances increase linearly after application of gate voltage, and drop slightly and then keep almost constant as the gate voltage was switched off, which is similar to the results shown in the main text, suggesting that oxygen vacancies were created at all different gas flow ratios. This result is different from that in the electrolyte gated STO, in which the oxygen vacancies were created only at vacuum, $N_2$ and Ar gas atmosphere, and the creation of oxygen vacancies was suppressed at pure oxygen atmosphere.[5] This indicates that oxygen vacancies were more easily to be created in gated NBCO, which could be seen from the fact the oxygen in CuO chains are prone to reduce from 7 to 6 under thermal treatment only at moderate temperature of 450 ~ 550 °C under low oxygen partial pressure,[6-7] while in STO higher temperature above 900 °C and higher vacuum were needed to create oxygen vacancies.[8] Therefore, even in the oxygen atmosphere, the oxygen vacancies are also created in gated

NBCO under electric field. For PCCO, however, at all different gas flow ratios, especially even at pure $O_2$ gas atmosphere, the resistances show similar behavior with time, the resistances decrease sharply as the gate voltage is applied and then tend to saturate with time, and almost recover to its initially value as the gate voltage is switched off, which is consistent with the result in the main text. This indicates that oxygen vacancies were not created during gating process, and that the gating effect in PCCO is mainly the electrostatic charging.

### 5. Control experiments of transport and ellipsometry measurements

In this section, we did control experiments of transport and ellipsometry measurements to support our discussions in main text.

First, we studied the deterioration effects on electrical transport properties incurred from the operations from RT3 to RT4 in Figure 2 in the main text. The operations from RT3 to RT4 included heating to 300 K, sample transfer out of PPMS chamber, acetone cleaning and blowing dry with nitrogen gas. We could identify that the main effects during these operations should come from the chemical reaction with ionic liquid above 240 K. Three samples were then prepared for the study. We put ionic liquid on sample 1 in air at 300 K for the chemical reaction for about 5 min and then removed the ionic liquid by acetone cleaning for about 2 min. The treatment on sample 2 was acetone cleaning alone for about 2 min. The measured RT curves before and after the treatments are shown in Figure S7a. We can see that the effect from acetone cleaning alone was negligibly small while the ionic liquid treatment plus acetone cleaning deteriorated the electrical transport properties significantly. Sample 1 changed from SC to insulating state. Thus, it is suggested that the deterioration effects incurred from operations from RT3 to RT4 was mainly caused by chemical reaction with ionic liquid above 240 K during warm up and sample taking out process. Sample 3 was first affected by the ionic liquid treatment in air at 300 K for a longer duration of 70 min and then followed by oxygen annealing treatment. The measured RT curves before and after these treatments are shown in Figure S7b. We can see that it changed from SC state to insulating state by the chemical reaction and the resistance could not be reduced further by subsequent oxygen annealing treatment. This excludes the possibility of oxygen vacancy formation from the chemical reaction with ionic liquid. That is, the deterioration effects at high temperature and in air cause the resistance after the removal of IL (RT4 in the main text) to be higher than the resistance at $V_g$ =1.6 V (RT3 in the main text). Therefore, if the deterioration effect is suppressed, oxygen vacancies formed during the electrolyte gating process will be filled after oxygen annealing, the electrical transport properties (RT5) will recover to the initial state (RT0).

To confirm that the changes of dielectric function and thus the oxygen vacancy formation in CuO chains are not caused by the chemical reaction with ionic liquid plus acetone cleaning, we did ellipsometry measurements in Figure S8a. We can see that the dielectric function almost did not change after the ionic liquid treatment for 6 min plus acetone cleaning for 2 min, in contrast to the huge enhancement of the 4.1 eV peak after electrolyte gating in Figure 3b in the main text. Hence, we learn that the chemical reaction with ionic liquid can deteriorate the electrical transport properties of NBCO film but this effect is not related to oxygen vacancy formation. Further study is needed to fully understand the chemical reaction.

## 6. Oxygen vacancy formation supported by XRD and AFM measurements

Figure S9 shows the reciprocal space mappings (RSM) data of 7-uc NBCO film before and after gating. The measurements were performed by high-resolution X-ray diffractometry (HR-XRD) in the X-ray demonstration and diffractometry (XDD) beamline at Singapore Synchrotron Light Source (SSLS). Figure S9 (a) and (b) show the (002) and (013) mapping for sample before gating. One can see that there is no tilt of NBCO to STO substrate and that the lattice of NBCO film is fully strained to the substrate. The $c$-axis and in-plane lattice parameters are 11.805 Å and 3.905 Å, respectively. Figure S9 (c) and (d) show the (002) and (013) mapping for sample after gating. The L value is lower than that before gating (from 2.3176 to 2.2986), indicating the elongation of the $c$-axis lattice parameter to be 11.902 Å. The lattice of NBCO film is still fully strained to the substrate after gating, so the in-plane lattice parameter is still 3.905 Å. In RE-123 cuprate system, as the oxygen content decreases, lattice parameter $c$ increases, in-plane parameter $a$ increases and $b$ decreases.[6-7] In the present study, since the NBCO film is ultrathin, the film is fully strained to STO substrate before and after gating. Therefore, the $c$-axis lattice parameter increases after gating while the in-plane parameter keeps unchanged. Overall, the significant increase in $c$ lattice parameter and therefore the expansion of the unit cell were observed after gating, indicating that oxygen vacancies were created in NBCO during the gating process.

We also measured the surface morphology of the NBCO film before and after gating by atomic-force microscopy (AFM), as shown in Figure S10. We can see that the structure became looser and a pit structure was produced after gating. These might reflect the channels for oxygen migration out of the sample film. The RMS roughness was increased from 1.03 nm to 1.73 nm. The looser structure implies expansion of the sample volume, which is consistent with the expansion of the NBCO unit cell after gating suggested by XRD measurements.

**Figures and Captions:**

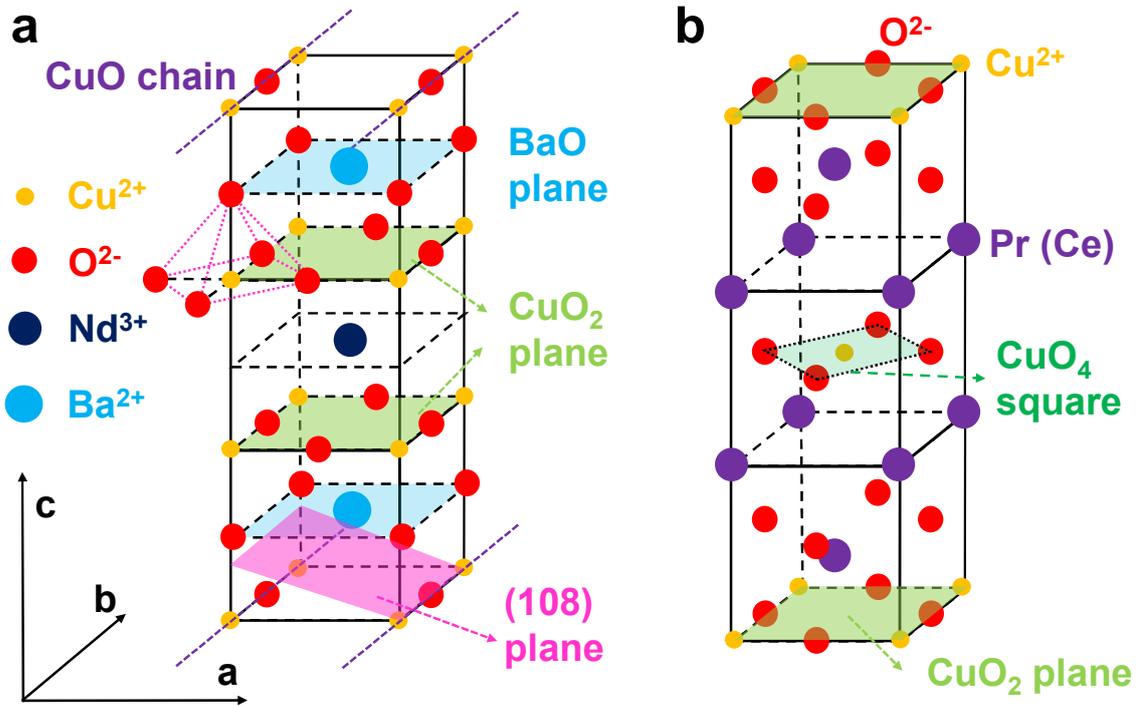

Figure S1. The unit cells of (a) $NdBa_2Cu_3O_{7-\delta}$ (NBCO) and (b) $Pr_{2-x}Ce_xCuO_4$ (PCCO).

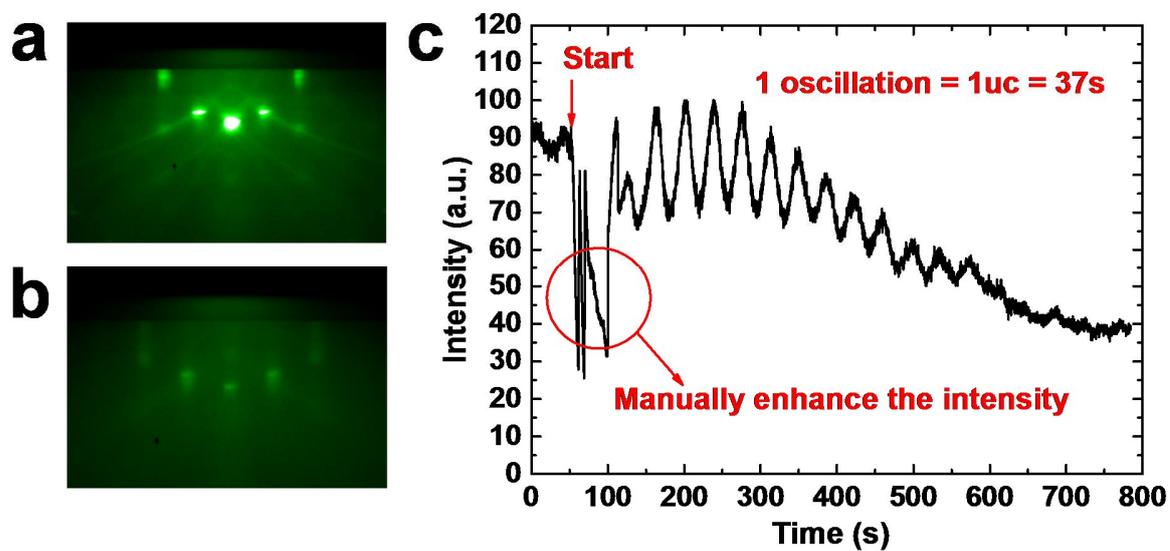

Figure S2. RHEED patterns (a) before and (b) after NBCO film growth on STO (100) substrate and (c) RHEED intensity oscillations during the initial 13-uc NBCO growth.

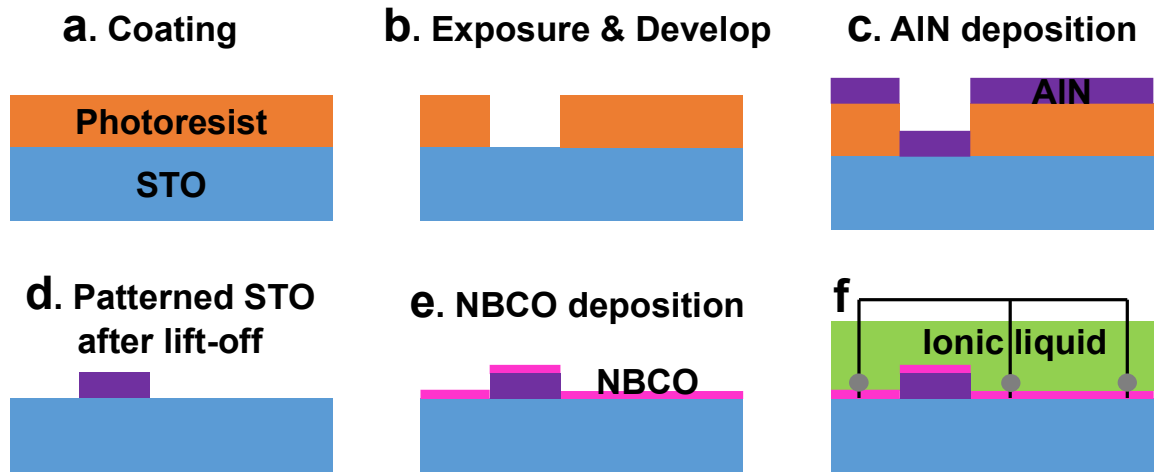

Figure S3. The schematic overview of the fabrication processes of electrolyte gating device.

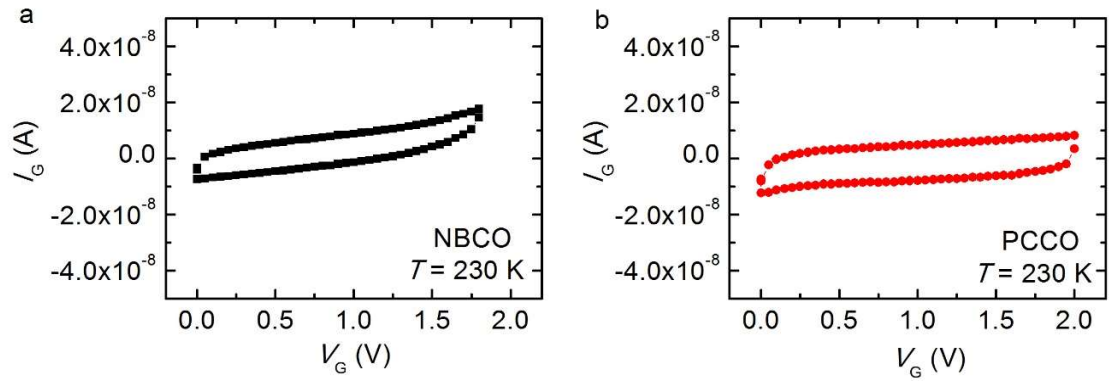

Figure S4. Gate voltage ($V_G$) dependence of the leakage currents ($I_G$) for (a) NBCO device and (b) PCCO device. The gating experiments were performed at 230 K.

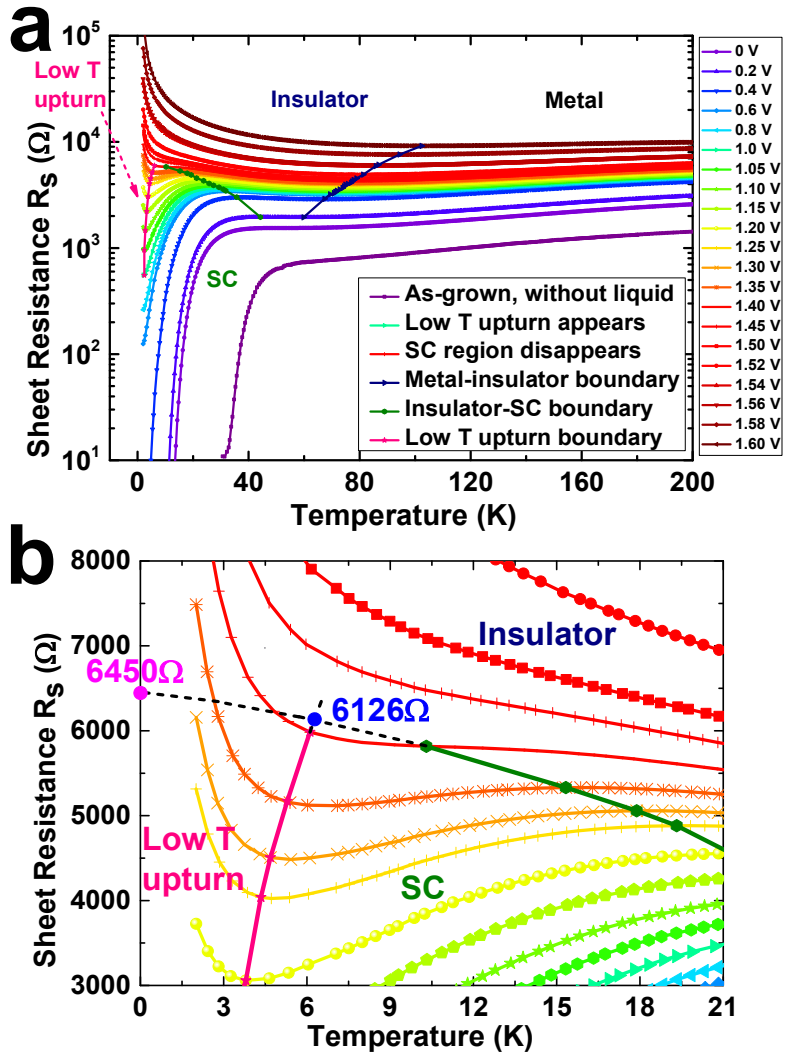

Figure S5. (a) Continuous modulation of SIT in NBCO by electrolyte gating. The data is divided into four electronic regions: metal, insulator, superconductor (SC), and low temperature ($T$) upturn. (b) Enlarged region for the insulator-SC and the low $T$ upturn boundaries.

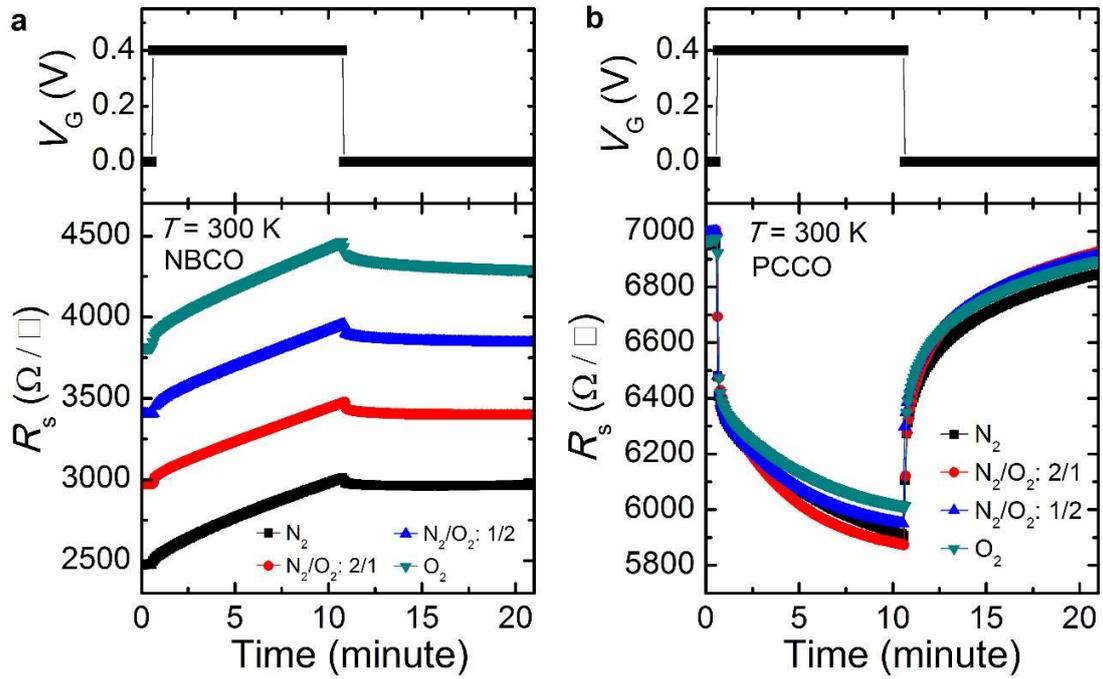

Figure S6. Sheet resistance $R_s$ as a function of time at various $N_2$ and $O_2$ gas flow ratio for (a) NBCO and (b) PCCO. The four gas flow ratios are pure $N_2$ gas, $N_2/O_2$ flow ratio of 2/1 and 1/2, and pure $O_2$ gas. At all different gas flow ratios, the gate voltage $V_G$ is initially set to be 0 V, and then 0.4 V for 10 minutes, and finally back to 0 V for 10 minutes. The measurements were performed at 300 K.

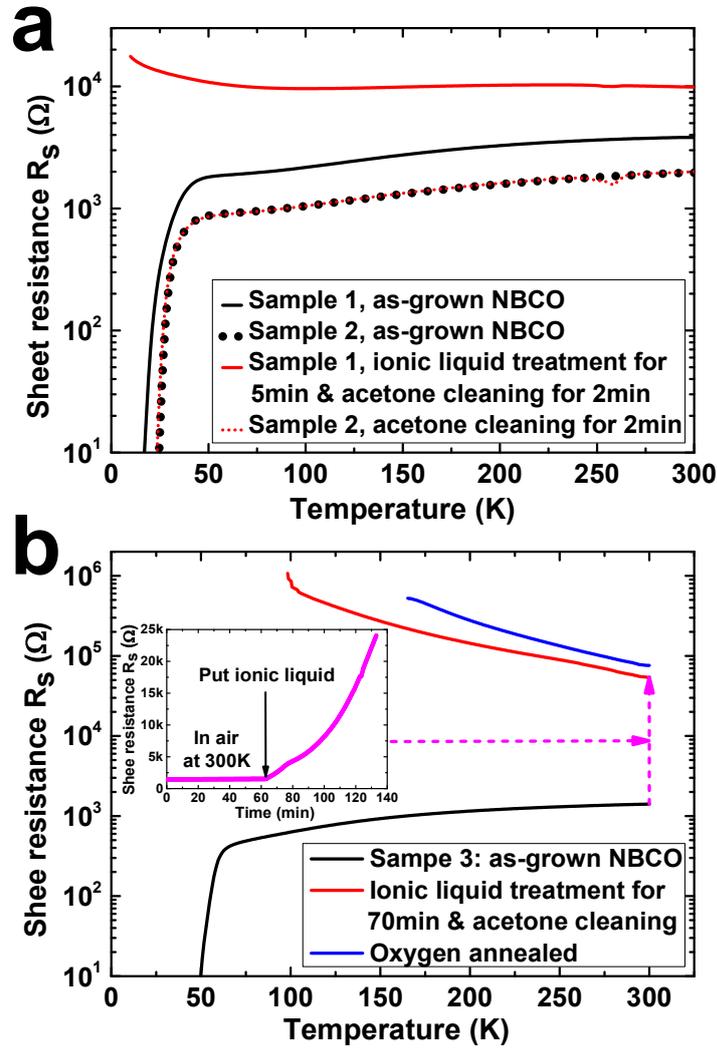

Figure S7. Control experiments of transport measurements of RT curves to study the deterioration effects on NBCO film due to the chemical reaction with ionic liquid. (a) Sample 1: the treatment was to mimic the operations from RT3 to RT4 (Figure 2a), including ionic liquid treatment in air at 300 K (chemical reaction) for about 5 min plus acetone cleaning for about 2 min. Sample 2: the treatment was acetone cleaning alone for about 2 min. (b) Sample 3: ionic liquid treatment in air at 300 K for a longer duration of 70 min, acetone cleaning for about 2 min, followed by oxygen annealing at 540 °C for 1 hour. Inset: time dependence of the $R_s$ before and after the ionic liquid was put on the sample surface in air at 300 K.

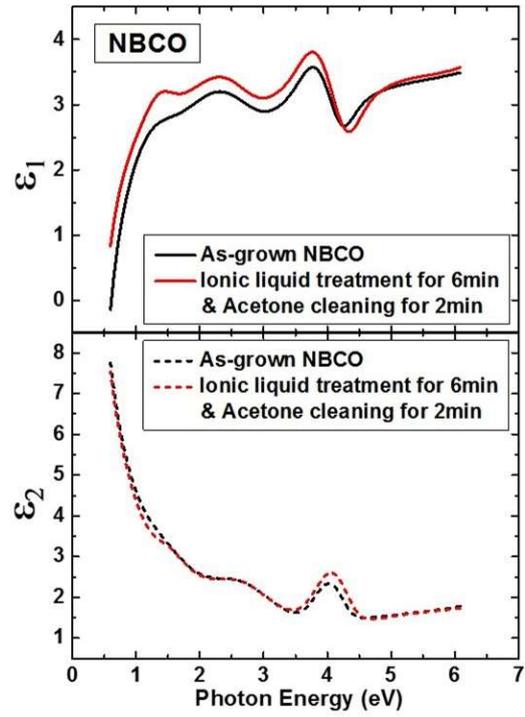

Figure S8. Control experiments of spectroscopic ellipsometry measurements of dielectric function $\varepsilon_1 + i\varepsilon_2$ for the as-grown NBCO film and the film with ionic liquid treated for 6 min and acetone cleaned for 2 min. Solid line is for real part $\varepsilon_1$ and dashed line is for imaginary part $\varepsilon_2$.

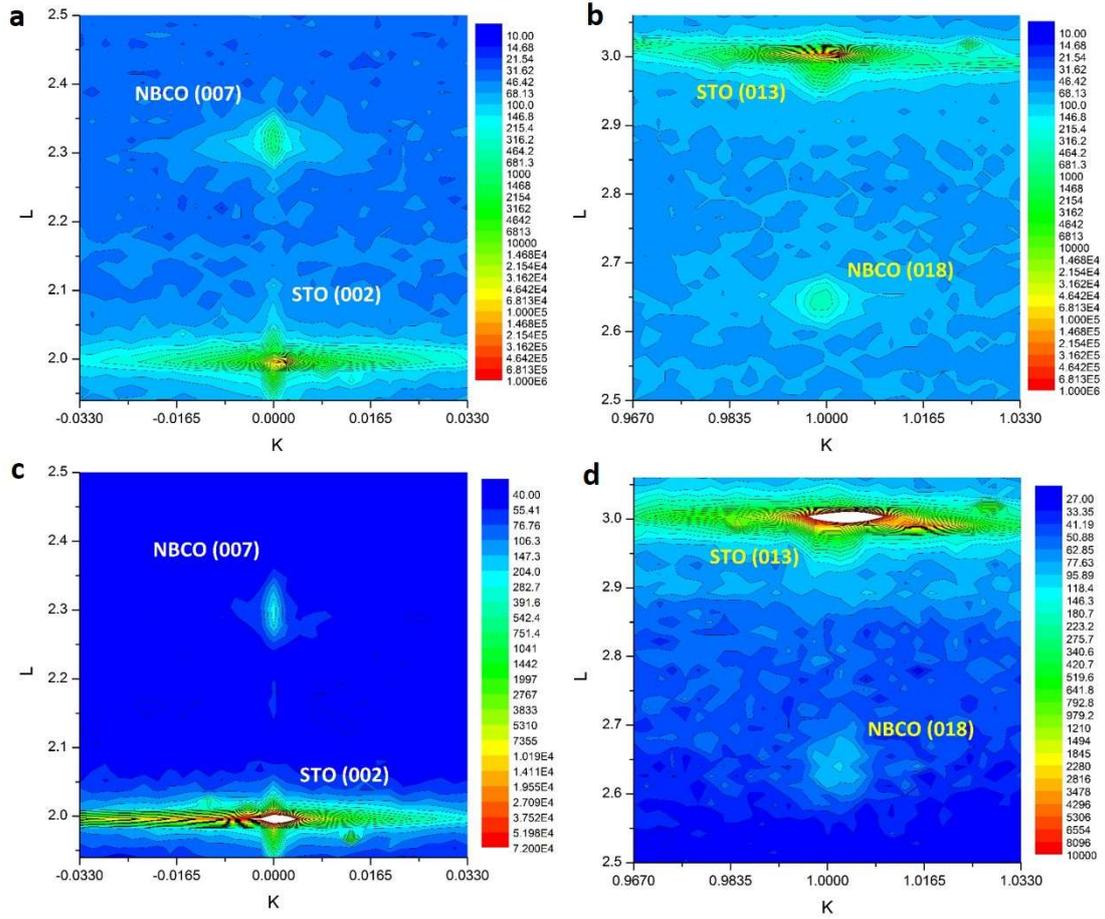

Figure S9. (a) and (b): RSMs of 7-uc NBCO before gating. (a) (002) mapping, showing that there is no tilt of NBCO film to the substrate STO; (b) (013) mapping, showing that NBCO (018) is located right down the substrate (013) around K=1, indicating that the lattice of NBCO film is fully strained to the substrate. (c) and (d): RSMs of NBCO after gating. (c) (002) mapping, showing that there is still no tilt of NBCO film to the substrate STO, but the L value is lower than that in (a) (from 2.3176 to 2.2986), indicating the elongation of the lattice parameter $c$ along the normal direction of substrate after gating; (d) (013) mapping, showing that NBCO (018) is still located right down the substrate (013) around K=1, indicating that the lattice of NBCO film is still fully strained to the substrate after gating.

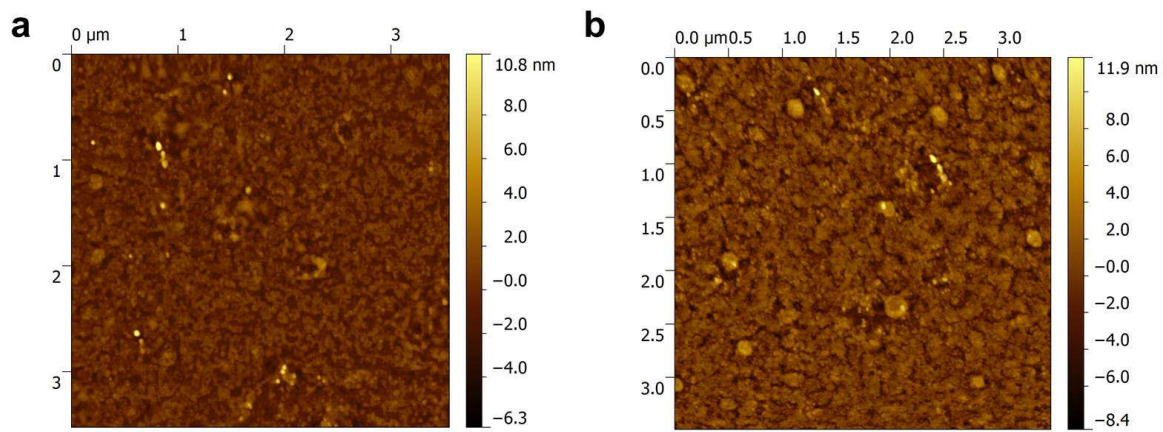

Figure S10. AFM surface morphology of the NBCO film (a) before gating (as-grown) and (b) after gating (electrolyte gated and removal of ionic liquid). The RMS roughness before gating is 1.03 nm and after gating is 1.73 nm.